\documentstyle[sprocl,psfig]{article}

\def\be{\begin{equation}}
\def\ee{\end{equation}}
\def\bea{\begin{eqnarray}}
\def\eea{\end{eqnarray}}
\def \apj {{\it Astrophys. J.\/}}

\def \aj  {{\it Astron. J.\/}}
\def \aap {{\it Astron. Astrophys.\/}}

\def\gtsim {>\kern-1.2em\lower1.1ex\hbox{$\sim$}~}   % Greater than sim
\def\ltsim {<\kern-1.2em\lower1.1ex\hbox{$\sim$}~}   % Less than sim

\begin{document}

%To Prof Nick Karayiannis -- do read this:-
%If needed the word of Chapter~1, you can type in at the 
%\title{}. The words will be in caps and lowercase. 
%For chapter title can be in all caps or in caps and lowercase.
%It is up to the author to type for the case sensitive but 
%all articles must be in the same style. 
%But mostly for Review Volume are without this Chapter~1.
%Thank you
%Jessie   13/4/2000

\title{COSMIC SUPERNOVA RATE HISTORY \\
AND \\ TYPE IA SUPERNOVA PROGENITORS}

\author{C. KOBAYASHI$^1$, K. NOMOTO$^1$, and T. TSUJIMOTO$^2$}

\address{$^1$ Department of Astronomy, School of Science,
University of Tokyo, Bunkyo-ku, Tokyo 113-0033, Japan \\
$~\,\, ^2$ National Astronomical Observatory, 
Mitaka, Tokyo 181-8588, Japan} 


\maketitle
\abstracts{
Adopting a single degenerate scenario for
Type Ia supernova progenitors with the metallicity effect,
we make a prediction of the cosmic supernova rate history 
as a composite of the supernova rates in spiral and elliptical galaxies,
and compare with the recent observational data up to $z \sim 0.55$.
}

\section{Metallicity Effects on Type Ia Supernova Progenitors
and Chemical Evolution of Galaxies}

We adopt a single-degenerate (SD) scenario, which assumes that
a C+O white dwarf accretes H-rich meterials from the binary companion star
and grows its mass to the Chandrasekhar mass to explode as a
Type Ia supernova (SN Ia).
There are two progenitor systems: One is a red-giant
(RG) companion with the initial mass of $M_{\rm RG,0} \sim 0.9-1.5 M_\odot$
and an orbital period of tens to hundreds days 
\cite{hkn96} \cite{hkn99} \cite{hknu99}.
The other is a near main-sequence (MS) companion with an initial mass of
$M_{\rm MS,0} \sim 1.8-2.6 M_\odot$ and a period of several tenths of a
day to several days \cite{li97} \cite{hknu99}.
Optically thick winds from the mass accreting WD play an
essential role in stabilizing the mass transfer and escaping from
forming a common envelope. 
Since the optically thick winds are driven by a
strong peak of iron lines, the occurrence of SNe Ia
depends strongly on the iron abundance.
If the iron abundance is as low as [Fe/H] $\ltsim -1.1$,
then the wind is too weak for SNe Ia to occur \cite{kob98}.

%The lifetimes of SNe Ia are determined from the
%main-sequence lifetimes of companion stars.
%The distribution function of companions is approximated by a
%power-law mass spectrum with a slope of $0.35$. 
%The fraction of primary stars of $3-8 M_\odot$ 
%which eventually produce SNe Ia is
%set to be $0.05$ for the MS+WD system and $0.02$ for the RG+WD system,
%which are adjusted to reproduce the chemical evolution 
%in the solar neighborhood.

%\section{Chemical Evolution in the Solar Neighborhood}

Our model successfully reproduces [O/Fe]-[Fe/H] relation
observed by oxygen forbidden lines,
which has a constant [O/Fe] of $\sim 0.45$ at [Fe/H] $\ltsim -1$
and a decrease in [O/Fe] with increasing metallicity.
If we do not include the metallicity effect, the largest companion
star with $M \sim 2.6 M_\odot$ produce SNe Ia at the age of $\sim 0.5$ Gyr
and decrease [O/Fe] too early to be compatible with the observations. 
For the metallicity dependent SD scenario, 
SNe Ia occur at [Fe/H] $\gtsim -1$,
which naturally reproduce the observed break in [O/Fe] at [Fe/H] $\sim -1$
\cite{kob98}.

%\section{Present Supernovae Rate in Galaxies}

We construct the galaxy models to meet the latest observational 
constraints of chemical and photometrical properties,
and can reproduce the present supernova rates in the galaxy.
For spirals, owing to the presence of the {\it two} kinds of 
the SN Ia progenitor systems (MS+WD and RG+WD) with 
shorter ($0.5-1.5$ Gyr) and longer lifetimes ($2-20$ Gyr), respectively,
we can explain the difference in the
relative ratio of the SN Ia to SN II rate ${\cal R}_{\rm Ia}/{\cal
R}_{\rm II}$ between early and late types of spirals.  
For ellipticals, owing to
the presence of the RG+WD systems with over $10$ Gyr lifetime, SNe Ia
can be seen even at present in ellipticals where the star formation
has already ceased more than 10 Gyr before 
because of a supernova-driven galactic wind \cite{kob99}.

\section{Cosmic Supernovae Rate}

Galaxies that are responsible for the cosmic SFR have different
timescales for the heavy-element enrichment, and the occurrence of
supernovae depends on the metallicity therein. Therefore
we calculate the cosmic supernova rate by summing up the supernova 
rates in spirals and ellipticals with the ratio of the relative mass
contribution \cite{kob99}.
Here we adopt
$H_0=65$ km s$^{-1}$ Mpc$^{-1}$, $\Omega_0=0.3$, $\lambda_0=0.7$, and
the formation epoch of $z_{\rm f}=5$.

Figure 1 shows the cosmic supernova rates in cluster galaxies.
The SN Ia rate in spirals drops at $z \sim 1.9$ 
because of the low-metallicity inhibition of SNe Ia. 
We can precisely test the metallicity effect 
by finding this drop of the SN Ia in spirals,
if high-redshift SNe Ia at $z \gtsim 1.5$ and their host galaxies
are observed with the Next Generation Space Telescope.
In ellipticals, the chemical enrichment takes place so early that 
the metallicity is large enough to produce SNe Ia at $z \gtsim 2$. 
The two peaks of SN Ia rates at $z \sim 2.6$ and $z \sim 1.6$ 
come from the MS+WD and the RG+WD systems, respectively.
The SN Ia rate in ellipticals decreases at $z \sim 2.6$,
which is determined from the shortest lifetime of SNe Ia of $\sim 0.5$ Gyr.
Thus, the total SN Ia rate decrease at the same redshift as ellipticals, i.e.,
$z \sim 2.6$.

\begin{figure}[ht]
\centerline{\psfig{figure=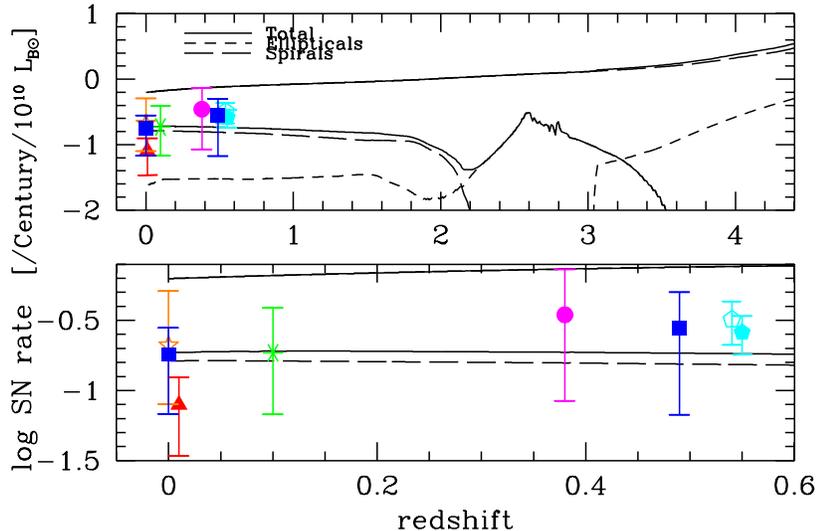,width=11.5cm}}
\caption[]{
The cosmic supernova rates (solid line) as the composite of ellipticals (short-dashed line) and spirals (long-dashed line).
The upper three lines show SN II rates, 
the lower three lines show SN Ia rates.
Observational data sources: 
triangle \cite{cap99}, star \cite{ham99}, asterisk \cite{har00},
circle \cite{pai96}, open pentagon \cite{pai99}, closed pentagon \cite{pai00}, 
and square \cite{rei00}}
\end{figure}

We also predict the cosmic supernova rates
assuming that the formation of ellipticals in field
took place for over the wide range of redshifts,
which is imprinted in the observed spectra of ellipticals
in the Hubble Deep Field. %\cite{fra98}.
The adopted SFRs are the same as the case of cluster galaxies,
but for the formation epochs $z_{\rm f}$ of ellipticals
distribute in the range of $0 \le z \le 5$.
Figure 2 shows the cosmic supernova rates in field galaxies.
As in Figure 1,
the SN Ia rate in spirals drops at $z \sim 1.9$. 
The averaged SN Ia rate in ellipticals decreases at $z \sim 2.2$ 
as a result of $\sim 0.5$ Gyr delay of the decrease in the SFR at $z \gtsim 3$.
Then, the total SN Ia rate decreases gradually
from $z \sim 2$ to $z \sim 3$.

Although the error bars are large, there is a hint that
the observed SN Ia rate decreases from $z \sim 0.4$ to the present.
If this is confirmed, it could imply that the rate of SNe Ia from
long-lived $0.9 M_\odot$ companions is lower than that assumed in our model.
From $z \sim 0.4$, the observed SN Ia rate seems to slightly decrease 
toward higher redshifts. 
To discuss their implications in terms of the progenitors' evolution,
we should exclude the luminosity uncertainties
in the unit of supernova rates
and introduce more detail galaxy models including internal structure
of a galaxy and galaxy number evolution.

The rate of SNe II in ellipticals evolves 
following the SFR without time delay.
Then, it is possible to observe SNe II in ellipticals around $z \sim 1$.
The difference in the SN II and Ia rates between
cluster and field ellipticals reflects the difference in the
galaxy formation histories in the different environments.

\begin{figure}[ht]
\centerline{\psfig{figure=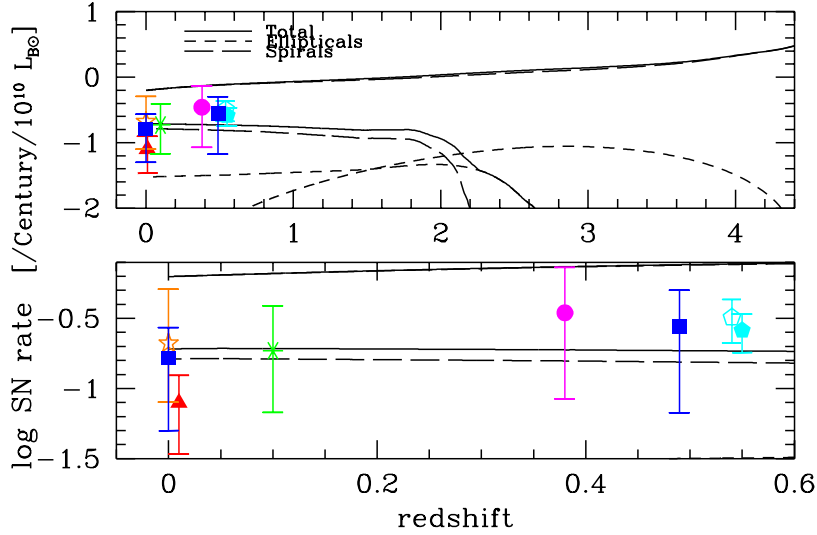,width=11.5cm}}
\caption[]{
The same as Figure 1, but for the formation epochs of ellipticals span at $1 \ltsim z \ltsim 4$, which corresponds to field ellipticals.
}
\end{figure}

\section{Discussion}

%We introduce a metallicity dependence of the SN Ia rate
%in galactic and cosmic chemical evolution models. 
%In our scenario involving a strong wind from WDs, 
%few SNe Ia occur at [Fe/H] $\ltsim -1$. 
%Our model successfully reproduces the observed chemical evolution
%in the solar neighborhood 
%and the present supernova rates in spirals and ellipticals.
We make the following predictions that can test this metallicity effect.
1) SNe Ia are not found in the low iron abundance environments
such as dwarf galaxies and the outskirts of spirals.
2) The SN Ia rate in spirals drops at $z \sim 2$ 
due to the low-iron abundance,
while SNe Ia can be found at $z \gtsim 2.5$ in cluster ellipticals, 
where the timescale of metal enrichment is sufficiently short.
3) If the formation of field ellipticals is protracted to lower redshifts,
the SNe Ia rate decreases from $z \sim 2$ in the field ellipticals.

Our predicted cosmic SN Ia rate is consistent with the recent
observation within error bars up to $z \sim 0.55$.
If stronger dependences on redshifts are found,
it would provide important constraints on the mass-range (i.e., lifetime)
of the WD's companions and metallicity effects.

%\section*{Acknowledgments}
%{\small This work has been supported in part by the grant-in-Aid for
%Scientific Research (08640336) and COE research (07CE2002).  
%C.K. thanks to the Japan Society for Promotion 
%of Science for a financial support. 
%We would like to thank I. Hachisu and M. Kato for providing us with their
%new results, and T. Kodama for providing us with the database of
%simple stellar population spectra.}

\end{document}